\documentclass[sigconf]{acmart}

\usepackage{xspace}
\usepackage{balance}
\usepackage[ruled]{algorithm2e}
\usepackage{graphics}
\usepackage{graphicx}
\usepackage{subfigure}
\usepackage{multirow}

\usepackage{url}

\newcommand{\ours}{AutoSeqRec\xspace}

\AtBeginDocument{%
  \providecommand\BibTeX{{%
    \normalfont B\kern-0.5em{\scshape i\kern-0.25em b}\kern-0.8em\TeX}}}

\copyrightyear{2023} \acmYear{2023}  \setcopyright{acmlicensed}\acmConference[CIKM '23]{Proceedings of the 32nd ACM International Conference on Information and Knowledge Management}{October   21--25, 2023}{Birmingham, United Kingdom}\acmBooktitle{Proceedings of the 32nd ACM International Conference on Information and Knowledge Management (CIKM '23), October   21--25, 2023, Birmingham, United Kingdom}\acmPrice{15.00}\acmDOI{10.1145/3583780.3614788}\acmISBN{979-8-4007-0124-5/23/10}

\begin{document}

\title{AutoSeqRec: Autoencoder for Efficient Sequential Recommendation}

\author{Sijia Liu}
\authornote{Both authors contributed equally to this research.}
\authornote{Also affiliated to Shanghai Key Laboratory of Data Science, China, and Shanghai Institute of Intelligent Electronics \& Systems, China.}
\email{sijialiu21@m.fudan.edu.cn}
\author{Jiahao Liu}
\authornotemark[1]
\authornotemark[2]
\email{jiahaoliu21@m.fudan.edu.cn}
\affiliation{%
  \institution{Fudan University}
  \city{Shanghai}
  \country{China}
}

\author{Hansu Gu}
\affiliation{%
  \city{Seattle}
  \country{United States}}
\email{hansug@acm.org}

\author{Dongsheng Li}
\affiliation{%
  \institution{Microsoft Research Asia}
  \city{Shanghai}
  \country{China}
}
\email{dongsli@microsoft.com}

\author{Tun Lu}
\authornote{Corresponding author.}
\authornotemark[2]
\affiliation{%
  \institution{Fudan University}
  \city{Shanghai}
  \country{China}
}
\email{lutun@fudan.edu.cn}

\author{Peng Zhang}
\authornotemark[2]
\affiliation{%
  \institution{Fudan University}
  \city{Shanghai}
  \country{China}
}
\email{zhangpeng\_@fudan.edu.cn}

\author{Ning Gu}
\authornotemark[2]
\affiliation{%
  \institution{Fudan University}
  \city{Shanghai}
  \country{China}
}
\email{ninggu@fudan.edu.cn}

\renewcommand{\shortauthors}{Sijia Liu et al.}
\begin{abstract}
Sequential recommendation demonstrates the capability to recommend items by modeling the sequential behavior of users. Traditional methods typically treat users as sequences of items, overlooking the collaborative relationships among them. Graph-based methods incorporate collaborative information by utilizing the user-item interaction graph. However, these methods sometimes face challenges in terms of time complexity and computational efficiency. To address these limitations, this paper presents \ours, an incremental recommendation model specifically designed for sequential recommendation tasks. \ours is based on autoencoders and consists of an encoder and three decoders within the autoencoder architecture. These components consider both the user-item interaction matrix and the rows and columns of the item transition matrix. The reconstruction of the user-item interaction matrix captures user long-term preferences through collaborative filtering. In addition, the rows and columns of the item transition matrix represent the item out-degree and in-degree hopping behavior, which allows for modeling the user's short-term interests.
When making incremental recommendations, only the input matrices need to be updated, without the need to update parameters, which makes \ours very efficient.
Comprehensive evaluations demonstrate that \ours outperforms existing methods in terms of accuracy, while showcasing its robustness and efficiency.
\end{abstract}



\maketitle
\section{Introduction}
Recommender systems have the ability to model user preferences and help users discover content of their interest from a vast amount of information~\cite{sarwar2001item,koren2009matrix,ning2011slim,steck2019embarrassingly,wang2019neural,he2020lightgcn,liu2023personalized}. Meanwhile, as user interests evolve over time, recommender systems are required to dynamically update and promptly adapt their recommendation strategies to keep up with the changes, capturing shifts in user interests in real time~\cite{kazemi2020representation,kumar2019predicting,zhang2021cope,liu2023triple,liu2022parameter,xia2022fire}. However,
training a recommendation model from scratch incurs significant costs, making it impractical to retrain the model every time there are interaction updates. Therefore, the sequential recommendation model faces a critical challenge of efficiently capturing evolving user interests. To address this challenge, two key problems need to be solved: 1) accurately capturing the evolution of user interests, and 2) swiftly adapting to changes through incremental updates.



Existing sequential recommendation models suffer from limitations in both aspects. On the one hand, traditional sequential methods~\cite{kumar2019predicting,wu2017recurrent,beutel2018latent} typically treat users as a sequences of items they have interacted with, overlooking the collaborative relationships among them.  Consequently, these methods fail to effectively leverage collaborative information, which is crucial for improving accuracy and personalization in recommendations. On the other hand, while some more recent graph-based approaches~\cite{chang2020continuous,zhang2021cope} can incorporate collaborative information by leveraging user-item interaction graphs, these methods often face challenges in terms of time complexity and computational efficiency. To address these limitations, we propose a novel recommendation model called \ours (\textbf{Auto}encoder for Efficient \textbf{Seq}uential \textbf{Rec}ommendation), which consists of the following two components.

To accurately capture the evolution of user interests, we propose a multi-encoder and decoder architecture. The first step involves constructing a user-item interaction matrix and an item transition matrix. These two matrices are then fed into a multi-information encoder, while three single-information decoders are employed for reconstruction purposes. The user-item interaction matrix contains the historical interactions between users and items. By reconstructing this matrix, \ours can effectively model collaborative information, providing a representation of users' long-term preferences. On the other hand, the item transition matrix captures the typical previous/next item that users interact with before/after the current item. Through reconstructing this matrix, \ours can capture users' real-time interactions, enabling a representation of their short-term interests. By combining collaborative information and item transition information, \ours is able to accurately capture the evolution of user interests.


To promptly adapt to changes through incremental updates, we introduce an efficient incremental inference process that incorporates new user-item interactions. This process begins by representing the single-hop item transition probability based on the reconstructed item transition matrix. It then efficiently approximates the multi-hop transition information by utilizing the compressed output of the multi-information encoder. When there are new updates on interactions, \ours only modifies the user-item interaction matrix and item transition matrix incrementally, while keeping the model parameters unchanged for inference. This approach enables \ours to efficiently adapt to new changes at the incremental level of a single interaction, facilitating real-time recommendation updates with minimal computational overhead.
The update process of \ours shares some insights with a contemporaneous work, IMCrrorect~\cite{liu2023recommendation}. However, unlike IMCrrorect, which focuses on recommendation unlearning tasks, \ours focuses on incremental recommendation tasks.



Our main contributions are as follows:
\begin{itemize}
\item We design a novel autoencoder architecture called \ours. By combining the user's long-term preference derived from collaborative information with the user's short-term interest modeled from item transition, our approach allows for a more accurate capture of user interest evolution and changes in behavior.
\item We develop an efficient incremental inference process that incorporates new user-item interactions without changing model parameters. It facilitates real-time recommendation updates at the incremental level of a single interaction.
\item Comprehensive evaluation on four datasets shows that \ours can significantly outperform existing sequential recommendation methods in  accuracy and efficiency. Moreover, \ours can be integrated with other sequential recommendation methods to  improve their performance.
\end{itemize}

\section{Preliminaries}
In this section, we introduce AutoRec~\cite{sedhain2015autorec}, a representative autoencoder framework for collaborative filtering.
Then, we describe the problem and discuss the limitations of the previous autoencoder-based method in solving this problem.

\subsection{Notations}
In this paper, we denote user set as $\mathcal{U}=\{u_1,u_2...,u_m\}$ and item set as $\mathcal{I}=\{i_1,i_2,...,i_n\}$, where $m$ is the number of users and $n$ is the number of items. 
Without loss of generality, we use $u$ to represent a user and $i$ to represent an item when their indices are not concerned.
The user-item rating matrix $Y\in\mathbb{R}^{m \times n}$ is partially observed.
Each user is represented by a partially observed vector $y^{(u)} = (Y_{u,1},...,Y_{u,n}) \in\mathbb{R}^n$, and each item is represented by a partially observed vector $y^{(i)} = (Y_{1,i},...,Y_{m,i}) \in\mathbb{R}^{m}$.

\subsection{AutoRec}

AutoRec~\cite{sedhain2015autorec} designs an item-based (or user-based) autoencoder which first takes as input each partially observed $y^{(i)}$ (or $y^{(u)}$), then projects it into a low-dimensional latent (hidden) space, and finally reconstruct $y^{(i)}$ ($y^{(u)}$) in the output space to predict missing ratings for static recommendation.

Taking item-based AutoRec as an example, which tries to solve the following reconstruction problem:
\begin{equation}
\min _\theta \sum_{\mathbf{y}^{(i)}}\|\mathbf{y}^{(i)}-h(\mathbf{\mathbf{y}^{(i)}} ; \theta)\|^2\text{,}
\end{equation}
where $h(\cdot)=f(g(\cdot))$ includes two processes: encoding and decoding.
The encoder $g(\cdot)$ and decoder $f(\cdot)$ are both composed of a single nonlinear layer.
Then, the parameters $\theta$ are learned using backpropagation.

Given learned parameters $\hat{\theta}$, the predicted rating of item-based AutoRec for user $u$ and item $i$ is
\begin{equation}
\hat{Y}_{u, i}=\left(h\left(\mathbf{y}^{(i)} ; \hat{\theta}\right)\right)_u.
\end{equation}

\subsection{Problem description}
Each user-item interaction can be represented by a tuple $(u^{(\tau)},i^{(\tau)})$, where $\tau=1,2,...$, indicates the sequential order in which the interaction occurred.
Suppose there are $N$ interactions in total, denoted as $\mathcal{S}^{(N)}=<(u^{(1)},i^{(1)}),(u^{(2)},i^{(2)}),...,(u^{(N)},i^{(N)})>$.
Now given a user $u$, we need to predict which item that user $u$ will interact with in the $N+1$-th interaction.
The output should be a $n$-dimensional vector, and each dimension represents the possibility score of interaction between user $u$ and the corresponding item.

AutoRec is a static recommendation method that cannot effectively utilize sequential information.
Although a lot of extensions have been proposed~\cite{liang2018variational,li2015deep,wu2016collaborative,yuan2019adversarial,steck2019embarrassingly,vanvcura2022scalable}, there is no prior work to extend AutoRec to incremental recommendation scenarios to the best of our knowledge.

\section{The Proposed Method}
In this section, we first introduce the overall architecture of \ours, and then present the implementation of incremental inference of \ours.

\subsection{Architecture}
\begin{figure*}[t]
  \centering
  \includegraphics[width=0.9\linewidth]{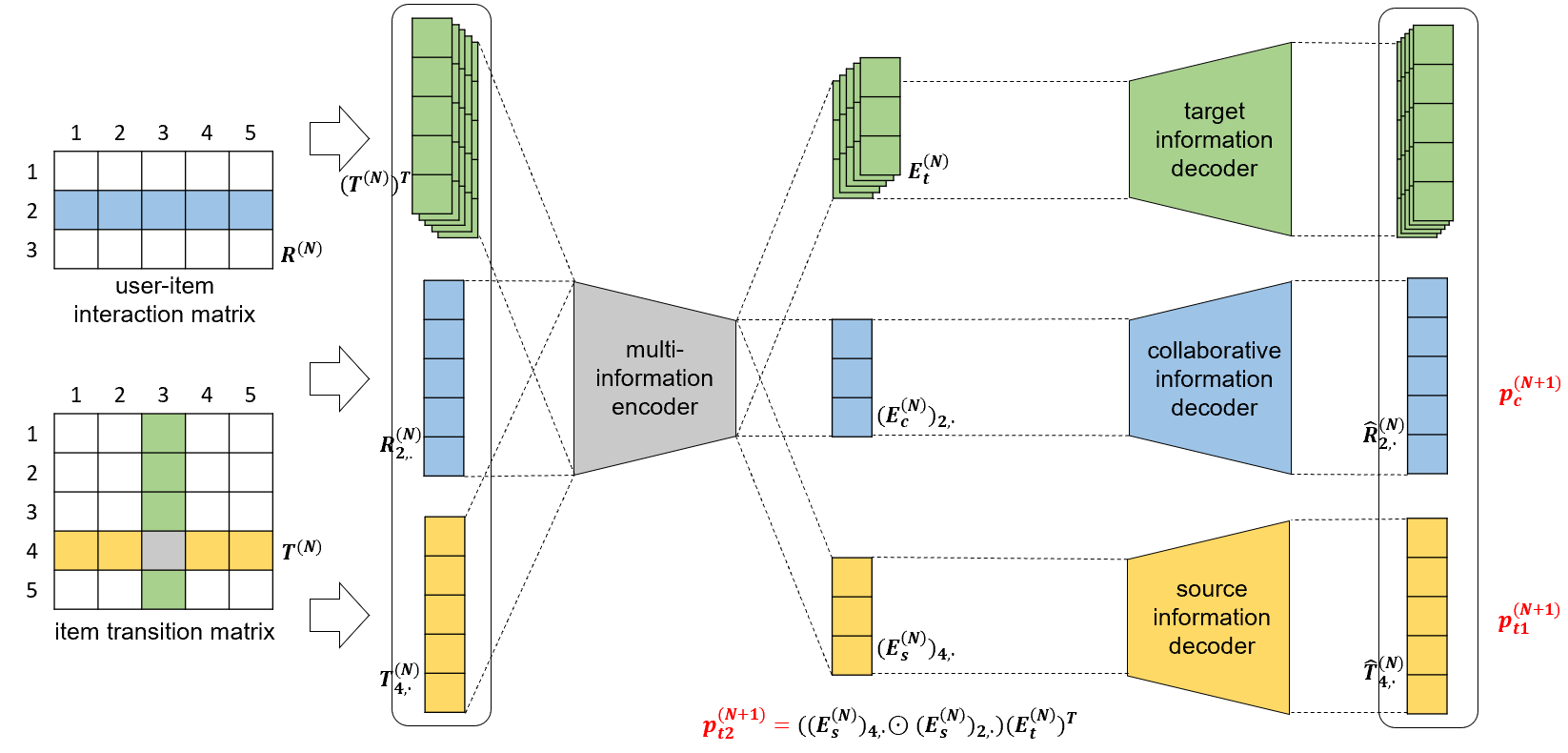}
  \caption{The inference process of \ours. Assuming that user $2$ last interacted with item $4$, this figure shows how \ours obtains the prediction score of user 2 interacting each item in the $N+1$-th interaction.}\label{fig:model}
  \Description{.}
\end{figure*}

In this subsection, we detail the \ours, including its implementation of the learning process and inference process.
The overall architecture of \ours is shown in Figure \ref{fig:model}.

\subsubsection{Learning Process} \label{learning}
We first construct the user-item interaction matrix $R^{(N)}\in\mathbb{R}^{m\times n}$ and item transition matrix $T^{(N)}\in\mathbb{R}^{n\times n}$, where the superscript indicates that there are a total of $N$ interactions.
The user-item interaction matrix $R^{(N)}$ stores the interaction records between users and items in $\mathcal{S}^{(N)}$, its entry at $u$-th row  and $i$-th column is defined as:
\begin{equation}
R^{(N)}_{u,i} = \begin{cases}
    1, & \text{user $u$ has interacted with item $i$ in $\mathcal{S}^{(N)}$,} \\
    0, & \text{otherwise.}
\end{cases}
\end{equation}
We define the interaction sequence of user $u$ in $\mathcal{S}^{(N)}$ as $\mathcal{S}_u^{(N)}=<i_u^{(1)},i_u^{(2)},...,i_u^{(n_u^{(N)})}>$, where $i_u^{(\tau)}$ is the item id of the $\tau$-th interaction of user $u$, $n_u^{(N)}$ is the number of interactions of user $u$ in $\mathcal{S}^{(N)}$.
The item transition matrix $T^{(N)}$ stores the transition information between item pairs, its entry at $i'$-th row and $i''$-th column is defined as:
\begin{equation}
T_{i',i''}^{(N)}=\sum_{u=1,...,m}\sum_{\substack{i_u^{(\tau)}=i',i_u^{(\tau+1)}=i''\\\tau=1,...,n_u^{(N)}-1}}1\text{.}
\end{equation}

\paragraph{Discussion about the input matrices.}
Each row of the user-item interaction matrix $R^{(N)}$ indicates which items the corresponding user has interacted with, which actually describes a user's personalized preference.
We call the elements related to $R^{(N)}$ {\bf collaborative information}.
Each row of the item transition matrix $T^{(N)}$ records the data of which item a user will interact with next time statistically, after the user interacts with the corresponding source item. Since it is statistical data from all users, this information is irrelevant to a specific user.
We call the elements related to $T^{(N)}$ {\bf source information}.
Correspondingly, each row of the transposition of the item transition matrix, $(T^{(N)})^\top$, records the data of which item the user interacted with last time if a user interacts with the corresponding target item. This is also statistical data from all users and does not contain personalized information.
We call the elements related to $(T^{(N)})^\top$ {\bf target information}.

In this paper, we implement the multi-information encoder (denoted as $f(\cdot)$) with a single nonlinear layer, more complex forms of the multi-information encoder are left for future work:
\begin{equation}
f(X)=\sigma(XW_e+\mathbf{b}_e)\text{,}
\end{equation}
where $X\in\mathbb{R}^{*\times n}$ is an input matrix with an unlimited number of rows, $W_e\in\mathbb{R}^{n\times k}$ is the weight matrix, $\mathbf{b}_e\in\mathbb{R}^{k}$ is the bias term, and $\sigma(\cdot)$ is the activation function. 
The user-item interaction matrix $R^{(N)}$, item transition matrix $T^{(N)}$ and its transpose $(T^{(N)})^\top$ are input into the multi-information encoder to obtain collaborative information embedding $E_c^{(N)}$, source information embedding $E_s^{(N)}$, and target information embedding $E_t^{(N)}$, respectively:
\begin{align}
E_c^{(N)}&=f(R^{(N)})\in\mathbb{R}^{m\times k}\text{,}\\
E_s^{(N)}&=f(T^{(N)})\in\mathbb{R}^{n\times k}\text{,}\\
E_t^{(N)}&=f((T^{(N)})^\top)\in\mathbb{R}^{n\times k}\text{.}
\end{align}

\paragraph{Discussion about multi-information encoder}
Firstly, the user-item interaction matrix, item transition matrix, and its transpose share consistent semantics. Each dimension represents a corresponding item, allowing them to be simultaneously compressed by an encoder.
Secondly, the multi-information encoder converts the input matrices into embedding representations with semantic meaning. This implies that if two users have similar interaction histories, i.e., their respective rows in the user-item interaction matrix are similar, their outputs after passing through the encoder will also exhibit similarity.
Likewise, if the transition patterns of two items (whether as source or target items) are comparable, indicating similar rows or columns in the item transition matrix, their outputs after going through the multi-information encoder will reflect this similarity.
Lastly, since the multi-information encoder incorporates collaborative information, source information, and target information simultaneously, the compressed representation generated by the multi-information encoder can effectively capture and consider all three types of information concurrently.

Next, we input collaborative information embedding $E_c^{(N)}$, source information embedding $E_s^{(N)}$, target information embedding $E_t^{(N)}$ into collaborative information decoder (denoted as $g_c(\cdot)$), source information decoder (denoted as $g_s(\cdot)$) and target information decoder (denoted as $g_t(\cdot)$) respectively, to reconstruct the user-item interaction matrix $R^{(N)}$, the item transition matrix $T^{(N)}$ and its transpose $(T^{(N)})^\top$:
\begin{align}
\tilde{R}^{(N)}&=g_c(E_c^{(N)})\in\mathbb{R}^{m\times n}\text{,}\\
\tilde{T}^{(N)}&=g_s(E_s^{(N)})\in\mathbb{R}^{n\times n}\text{,}\\
(\tilde{T}^{(N)})^\top&=g_t(E_t^{(N)})\in\mathbb{R}^{n\times n}\text{.}
\end{align}
Again, we only adopt the simplest setting, where each of the three decoders consists of a single nonlinear layer:
\begin{align}
g_c(X)&=\sigma(XW_c+\mathbf{b}_c)\text{,}\\
g_s(X)&=\sigma(XW_s+\mathbf{b}_s)\text{,}\\
g_t(X)&=\sigma(XW_t+\mathbf{b}_t)\text{,}
\end{align}
where $W_c,W_s,W_t\in\mathbb{R}^{k\times n}$ are the weight matrices, $\mathbf{b}_c,\mathbf{b}_s,\mathbf{b}_t\in\mathbb{R}^{n}$ are the bias terms.

Finally, we train \ours by reducing the reconstruction loss, using the Frobenius norm as the measure of the differences between the input matrices and the reconstructed matrices.
Specifically, we denote the reconstruction loss of the collaborative information decoder as $\mathcal{L}_c$, denote the reconstruction loss of the source information decoder as $\mathcal{L}_s$, and denote the reconstruction loss of the target information decoder as $\mathcal{L}_t$, which are defined as follows, respectively:
\begin{align}
\mathcal{L}_c&=||R^{(N)}-\tilde{R}^{(N)}||_F\text{,}\\
\mathcal{L}_s&=||T^{(N)}-\tilde{T}^{(N)}||_F\text{,}\\
\mathcal{L}_t&=||(T^{(N)})^\top-(\tilde{T}^{(N)})^\top||_F\text{.}
\end{align}
By minimizing the reconstruction loss, we aim to enable the decoders to recover the original data as accurately as possible.
The final loss function consists of three kinds of losses:
\begin{equation}
\mathcal{L}=\mathcal{L}_c+\mathcal{L}_t+\mathcal{L}_s\text{.}
\end{equation}
Through backpropagation during training, we iteratively update model parameters to minimize reconstruction loss. Since the reconstructions of three types of information are equally important, we do not need to set any hyperparameters to balance among them.


\subsubsection{Inference Process} \label{inference}
As aforementioned, the user-item interaction matrix reflects user personalized information.
In static scenarios, we can directly use the reconstructed user-item interaction matrix $\tilde{R}^{(N)}$ as the prediction result.
The collaborative information-based probability score of user $u$ interacting with all items in the $N+1$-th interaction can be expressed as:
\begin{equation}
\mathbf{p}_c^{(N+1)}=\tilde{R}^{(N)}_{u,\cdot}\in\mathbb{R}^n\text{.}
\end{equation}
Each dimension of $\mathbf{p}_c$ corresponds to an item.

However, in dynamic scenarios, the reconstructed interaction matrix $\tilde{R}^{(N)}$ can only reflect users' long-term preferences and cannot accurately reflect users' temporal interests.
To capture users' short-term interests, we further utilize item transition matrix.
For user $u$, whose last interacted item is $i_u^{(n_u^{(N)})}$ (for the convenience of narration, we use $i$ to refer to $i_u^{(n_u^{(N)})}$), we take the $i$-th row of the reconstructed item transition matrix as the short-term interest of user $u$.
Then the single-hop transition probability score of user $u$ is:
\begin{equation}
\mathbf{p}_{t1}^{(N+1)}=\tilde{T}^{(N)}_{i,\cdot}\in\mathbb{R}^n\text{.}
\end{equation}
Each dimension of $\mathbf{p}_{t1}$ corresponds to an item.

\begin{figure}[t]
\centering
\includegraphics[width=0.85\linewidth]{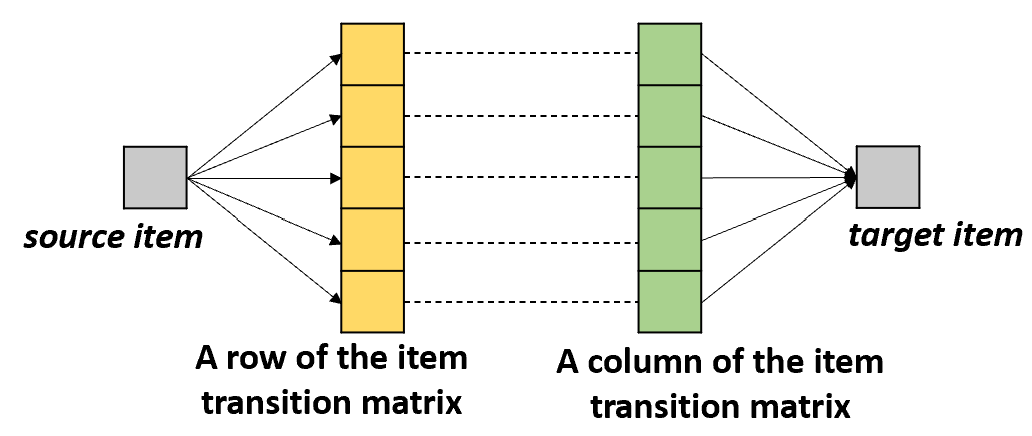}
\caption{Explanation of two-hops transition. A row of the item transition matrix records the data of which item a user will interact with next time, after the user interacts with the source item. A column of the item transition matrix records the data of which item the user interacted with last time if a user interacts with the target item. The inner product of the two vectors represents the probability score of the source item reaching the target item after two hops.}\label{fig:tran}
  \Description{}
\end{figure}

$\mathbf{p}_{t1}^{(N+1)}$ only considers the information of single-hop transition.
In order to more comprehensively consider the information of multi-hop transitions, a simple solution is to multiply $\mathbf{p}_{t1}^{(N+1)}$ and the item transition matrix $\tilde{T}^{(N)}$ continuously.
As shown in the Figure \ref{fig:tran}, $\mathbf{p}_{t1}^{(N+1)}$ reflects the probability score of the items that will be reached from a source item, and each column of the item transition matrix reflects the probability score of the items that a target item may come from.
So one multiplication can represent the probability score of reaching the target item from the source item after two hops.
However, the time complexity of this approach can be very high when the number of items is large.

We took a more efficient approach. As mentioned before, similar inputs will have similar outputs after passing through the multi-information encoder, which in turn will have similar decoder outputs. So instead of operating on the reconstruction matrix output by the decoder, we can operate on the embedded representation output by the multi-information encoder.
Specifically, we perform a Hadamard product operation on the collaborative information embedding of user $u$ ($(E_c^{(N)})_{u,\cdot}$) and $(E_s^{(N)})_{i,\cdot}$ to obtain a personalized transition embedding:
\begin{equation}
\tilde{\mathbf{p}}_{t2}^{(N+1)}=(E_s^{(N)})_{i,\cdot}\odot(E_c^{(N)})_{u,\cdot}\label{con:Hadamard}
\end{equation}
Furthermore, we multiply the $\tilde{\mathbf{p}}_{t2}^{(N+1)}$ by the target information embedding $E_t^{(N)}$ to obtain the two-hop transition information:
\begin{equation}
\mathbf{p}_{t2}^{(N+1)}=\tilde{\mathbf{p}}_{t2}^{(N+1)}(E_t^{(N)})^\top\in\mathbb{R}^n\text{.}\label{con:multiply}
\end{equation}
The multi-hop transition information can be obtained by multiplying the transpose of the target information embedding, $(E_t^{(N)})^\top$, multiple times.

Finally, we take the weighted sum of three types of probability scores (collaborative information probability score $\mathbf{p}_c^{(N+1)}$, one-hop transition probability score $\mathbf{p}_{t1}^{(N+1)}$, and two-hops transition probability score $\mathbf{p}_{t2}^{(N+1)}$) as the final prediction result, as follows:
\begin{equation}
\mathbf{p}^{(N+1)}=\lambda_1\mathbf{p}_c^{(N+1)}+\lambda_2\mathbf{p}_{t1}^{(N+1)}+(1-\lambda_1-\lambda_2)\mathbf{p}_{t2}^{(N+1)}\text{.}\label{con:coefficients}
\end{equation}
The coefficients $\lambda_1$ and $\lambda_2$ control the importance of different sources of information in the final prediction results.

\subsection{Incremental Inference}
In this subsection, we introduce how \ours effectively handles dynamic data and updates recommendation results in real time.
Specifically, we first analyze why the autoencoder-based CF model can naturally achieve incremental recommendation
, and then we give the concrete implementation of incremental recommendation of \ours.

\subsubsection{Analysis}
Representation learning-based recommendation methods, e.g., matrix factorization~\cite{koren2009matrix}, first obtain the embeddings of users and items, and then perform personalized ranking of items by calculating the similarity (e.g., inner product) between users and items.
Different from these representation learning-based methods, the autoencoder-based collaborative filtering method is a process of obtaining predicted interactions directly from observed interactions.
For example, AutoRec directly inputs the user-item interaction matrix into the autoencoder, and takes the reconstructed user-item interaction matrix as the prediction result.
Autoencoder-based methods are often used to solve the problem of static recommendation.

Actually, once an autoencoder is trained, it is equipped to handle various user interests.
Now, let us focus on the dynamic recommendation process for two users $a$ and $b$.
Assume that user $a$ likes to watch comedy movies in the early days, and user $b$ likes to watch science fiction movies during the same period.
Because the historical interaction records of the two users are different ($a$ interacts with more comedy movies, $b$ interacts with more science fiction movies), when a recommendation model based on autoencoder is built, $a$ will be recommended comedy movies, $b$ will be recommended science fiction movies.
Both user $a$ and user $b$ can obtain the desired personalized recommendation in that period, and the root cause of their different recommendation results is the difference in historical interaction (the difference between the corresponding rows of the interaction matrix).
If in the later stage, $a$'s interest has changed to science fiction movies.
At this time, the interaction record of $a$ (interacting more science fiction movies instead of comedy movies) is input into the recommendation model, and $a$ will be recommended science fiction movies.
The autoencoder-based model has this ability because it has learned how to recommend science fiction movies to users (e.g., user $b$) from the training data.
To sum up, the collaborative filtering method based on autoencoder actually supports incremental recommendation naturally. We only need to change the input matrix, and the weight of autoencoder does not need to be updated during the incremental inference process.

\subsubsection{Implementation}
We believe that \ours has the ability to handle various user interests and encode the historical interactions of users at different moments.
Therefore, we only need to change the input user-item interaction matrix $R$ and item transition matrix $T$, and then follow the same inference process to make predictions.
Specifically, assuming user $u$ interacts with item $i$ at $N+1$-th interaction, we can use the following steps to predict the next interaction: 
1) update the user-item interaction matrix $R^{(N+1)}$ and item transition matrix $T^{(N+1)}$ to contain the latest interaction state between user $u$ and item $i$.
2) input the updated user-item interaction matrix and item transition matrix into the \ours to obtain the recommendation result.
After the model training is completed, we can make accurate next-step interaction predictions based on the existing interaction history.
In addition, we can execute the above prediction process as many times as needed to achieve continuous prediction of the user's next interaction item.
This flexibility and efficiency enable \ours to quickly respond to user needs and provide personalized and high-quality recommendation services in practical applications.

\section{Experiments}
\begin{table}[t]\small
\centering
\caption{The statistics of the datasets used in the experiments.}
\label{table.statistics.item}
\begin{tabular}{c|cccc}
\hline
        & \multicolumn{1}{l}{\# Users} & \multicolumn{1}{l}{\# Items} & \multicolumn{1}{l}{\# Interactions} & \multicolumn{1}{l}{\# Unique Times} \\ \hline
ML-100K & 943                          & 1,349                        & 99,287                              & 49,119                              \\
ML-1M   & 6,040                        & 3,416                        & 999,661                             & 458,254                             \\
Garden   & 1,686                        & 962                        & 13,272                             & 1,888                             \\
Video   & 5,130                        & 1,685                        & 37,126                              & 1,946                               \\
\hline
\end{tabular}
\end{table}
In this section, we conduct comprehensive evaluation to answer the following research questions (RQs):
\begin{itemize}
    \item \textbf{RQ1} Does \ours outperform state-of-the-art methods in future interaction prediction task on dynamic graphs?
    \item \textbf{RQ2} How do different components of \ours affect model performance?
    \item \textbf{RQ3} Does the desgin of \ours help to reduce running time?
    \item \textbf{RQ4} Does the hidden embedding size and coefficients $\lambda_1$, $\lambda_2$ affect predictions?
    \item \textbf{RQ5} Can \ours be seamlessly integrated with other sequential recommendation models to improve their performance? 
\end{itemize}
\subsection{Experimental Setup}
\subsubsection{Dataset}
We use four publicly available datasets to evaluate the performance of \ours. MovieLens ~\cite{harper2015movielens}
is a movie rating dataset that is widely used in dynamic graph tasks. We use two versions of MovieLens: ML-100K and ML-1M. ML-100K contains 100K interactions collected during a seven-month period from September 19th, 1997 to April 22nd, 1998. ML-1M contains 1M interactions from users who joined MovieLens in 2000. Amazon~\cite{he2016ups} contains ratings on the e-commerce platform Amazon.com, and we adopt two sub-datasets, Garden and Video. We follow the same preprocessing pipeline in previous works~\cite{kang2018self,tang2018personalized} that users and items with fewer than 5 interactions are discarded. The statistical information of datasets is shown in Table ~\ref{table.statistics.item}. 
\subsubsection{Setting}
We run all the experiments on a server equipped with one NVIDIA TESLA T4 GPU and Intel(R) Xeon(R) Gold 5218R CPU. All the code of this work is implemented with Python 3.7.4~\footnote{The codes are available at \url{https://github.com/sliu675/AutoSeqRec}}. Three input matrices including user-item interaction matrix, item transition matrix and the transposition of item transition matrix are first divided into mini-batches separately, and then we union these mini-batches and randomly shuffle them. In the training process, we use the corresponding loss function according to which matrix is used in the mini-batches union. The hyper-parameters are tuned via grid search. More specifically, we search autoencoder's hidden dimension size in [32, 64, 128, 256] and coefficients $\lambda_1$, $\lambda_2$ from values in [0.0, 0.1, 0.2, $\cdots$, 1.0]. 

\begin{table*}[t]\small
    \caption{Accuracy comparison on the future interaction prediction task.}
    \label{tab:main results}
\begin{tabular}{l|c|c|c|c|c|c|c|c}
\hline\multirow{2}{*}{\text { Method }} & \multicolumn{2}{|c|}{\text { ML-100K }} & \multicolumn{2}{c|}{\text { ML-1M }} & \multicolumn{2}{c|}{\text { Video }} & \multicolumn{2}{c}{\text { Garden }} \\ 
\cline{2-9}
& \text { MRR } & \text { Recall@10 } & \text { MRR } & \text { Recall@10 } & \text { MRR } & \text { Recall@10 } & \text { MRR } & \text { Recall@10 } \\
\hline \text { LightGCN } & 0.012 & 0.025 & 0.014 & 0.031 & 0.019 & 0.036 & 0.025 & 0.087 \\
\text { RRN } & 0.042 & 0.084 & 0.038 & 0.071 & 0.039 & 0.074 & 0.079 & 0.169 \\
\text { JODIE } & 0.046 & 0.095 & 0.037 & 0.070 & 0.044 & 0.081 & 0.059 & 0.142 \\
\text { CoPE } & 0.049 & 0.111 & 0.040 & 0.085 & 0.036 & 0.078 & 0.082 & 0.176 \\
\text { FreeGEM } & 0.040 & 0.079 & 0.040 & 0.077 & 0.058 & 0.086 & 0.100 & 0.179 \\
\textbf { \ours } & \textbf{0.068} & \textbf{0.150} & \textbf{0.060} & \textbf{0.130} & \textbf{0.064} & \textbf{0.107} & \textbf{0.102} & \textbf{0.227} \\
\hline
\text { Rel. Imp.} & 38.78\% & 35.14\% & 50.00\% & 52.94\% & 10.34\% & 10.62\% & 2.00\% & 26.82\% \\
\hline
\end{tabular}
\end{table*}

\begin{table*}[t]\small
    \caption{Ablation study on the encoder hidden outputs operations.}
    \label{tab:ablation2}
\begin{tabular}{l|c|c|c|c|c|c|c|c}
\hline \multirow{2}{*}{\text {Setting}} & \multicolumn{2}{|c|}{\text { ML-100K }} & \multicolumn{2}{c|}{\text { ML-1M }} & \multicolumn{2}{c|}{\text { Video }} & \multicolumn{2}{c}{\text { Garden }} \\
\cline{2-9}
& \text { MRR } & \text { Recall@10 } & \text { MRR } & \text { Recall@10 } & \text { MRR } & \text { Recall@10 } & \text { MRR } & \text { Recall@10 } \\
\hline \text { $b \cdot{c}$ } & 0.050 & 0.113 & 0.031 & 0.064 & 0.016 & 0.026 & 0.058 & 0.123 \\
\text { $a \cdot{c}$ } & 0.038 & 0.079 & 0.020 & 0.037 & 0.030 & 0.060 & 0.086 & 0.203 \\
\textbf { $(a \odot{b})\cdot{c}$ } & \textbf{0.058} & \textbf{0.130} & \textbf{0.037} & \textbf{0.076} & \textbf{0.040} & \textbf{0.086} & \textbf{0.097} & \textbf{0.206} \\
\hline
\end{tabular}
\end{table*}

\begin{table*}[t]\small
    \caption{Ablation study on the decoder outputs of user-item interaction matrix and item transition matrix.}
    \label{tab:ablation1}
\begin{tabular}{l|c|c|c|c|c|c|c|c}
\hline \multirow{2}{*}{\text {Setting}} & \multicolumn{2}{|c|}{\text { ML-100K }} & \multicolumn{2}{c|}{\text { ML-1M }} & \multicolumn{2}{c|}{\text { Video }} & \multicolumn{2}{c}{\text { Garden }} \\
\cline{2-9}
& \text { MRR } & \text { Recall@10 } & \text { MRR } & \text { Recall@10 } & \text { MRR } & \text { Recall@10 } & \text { MRR } & \text { Recall@10 } \\
\hline \text {only  hidden} & 0.058 & 0.130 & 0.037 & 0.076 & 0.040 & 0.086 & 0.097 & 0.206 \\
\text {hidden $+$ interaction matrix} & 0.062 & 0.138 & 0.040 & 0.081 & 0.038 & 0.078 & 0.091 & 0.215 \\
\text {hidden $+$ transition matrix} & 0.060 & 0.132 & 0.058 & 0.126 & 0.062 & 0.106 & 0.101 & 0.226 \\
\textbf {all} & \textbf{0.068} & \textbf{0.150} & \textbf{0.060} & \textbf{0.130} & \textbf{0.064} & \textbf{0.107} & \textbf{0.102} & \textbf{0.227} \\
\hline
\end{tabular}
\end{table*}

\subsubsection{Baseline Models}
We compare \ours with the following state-of-the-art interaction graph modelling methods:
\begin{itemize}
    \item LightGCN~\cite{he2020lightgcn} is a representative collaborative filtering algorithm based on GNNs, which ignores time information.
    \item RRN~\cite{wu2017recurrent} models user and item interaction sequences with separate RNNs for sequential recommendation.
    \item JODIE~\cite{kumar2019predicting} also generates node embeddings using RNNs, and can also estimate user embeddings trajectories.
    \item CoPE~\cite{zhang2021cope} uses an ordinary differential equation-based GNN to model the evolution of network.
    \item FreeGEM~\cite{liu2022parameter} is a state-of-the-art parameter-free method for dynamic graph learning tasks.
\end{itemize}

\subsection{Future interaction prediction (RQ1)}

The future prediction prediction task is defined as follows: given all interactions till time $t$, we predict which item will user $u$ interact with at time $t$? We use the same data splitting as JODIE's~\cite{kumar2019predicting} and CoPE’s~\cite{zhang2021cope} to separate the dataset by time. Specifically, we use the earliest 80\% interactions for training, the following 10\% data for validation, and the latest 10\% data for test. We adopt mean reciprocal rank (MRR) and Recall@10 to evaluate the performance of future interaction prediction, in which MRR is the average of the reciprocal rank and Recall@10 is the fraction of interactions in which the ground truth items are within the top 10 recommendations,  following JODIE, CoPE and FreeGEM. 

Table \ref{tab:main results} shows the results of \ours compared with the above five state-of-the-art methods. 
We observe that \ours significantly outperforms all compared methods in all datasets across both metrics Recall@10 and MRR on the four datasets. We calculate the relative improvement (Rel. Imp.) of \ours over the baseline as (performance of \ours~$-$ performance of baseline)/(performance of baseline). Across all datasets, the minimum improvement of \ours is 2.00\% - 50.00\% in terms of MRR and 10.62\% - 52.94\% in terms of Recall@10. We notice that LightGCN performs the worst on most datasets.
This is largely due to the fact that we split the dataset by time according to this experiment setting.
Since \ours makes full use of user-item collaborative information and one-hop or even two-hops of item transition information, \ours outperforms the other methods. 

\subsection{Ablation studies (RQ2)}

In this section, we first conduct ablation study on the operations applied to the three types of information from the encoder. 
Then, we test whether the decoder outputs of user-item interaction matrix and item transition matrix are helpful or not. 
\subsubsection{Encoder hidden outputs operations}
This experiment test the effectiveness of 1) the Hadamard product operation on the collaborative information embedding and source information embedding and 2) the inner product operation between the personalized transition embedding and the target information embedding. 
We use a, b, c to represent the collaborative information embedding of user $u ((E_c)_{u,\cdot})$, $u$'s last interacted item $i$'s source information embedding $((E_s)_{i,\cdot})$ and the target information embedding $E_t$. We observe that combining a, b and c obtain the best performance among all the combinations as shown in Table \ref{tab:ablation2}. Moreover, operation $a \cdot c$ performs better than $b \cdot c$ on Amazon datasets, while the opposite is true on MovieLens dataset. One possible reason is that the two Amazon datasets used in our experiments are more sparse than the MovieLens datasets, which may lead to less robust statistics in the item transition matrix. Therefore, the collaborative information from the user-item interaction matrix becomes more important.

\begin{table*}[t]\small
\centering
\caption{Running time comparison on the four datasets.}
\label{tab:running}
\begin{tabular}{@{}c|cccc@{}}
\hline
                     & ML-100K & ML-1M      & Video   & Garden \\ \hline
JODIE                 & 8h47min56s (298.8x)       & 374h7min53s (1989.5x)         & 46min18s (38.6x)       & 16min51s (16.3x)     \\
CoPE                 & 16h51min5s (572.3x)       & 1025h24min6s (5452.7x)          & 1h38min33s (82.1x)      & 1h6min52s (64.7x)     \\
FreeGEM              & 3min12s (1.8x) & 5h54min18s (31.4x) & 2min46s (2.3x) & 1min7s (1.1x) \\
\ours & 1min46s & 11min17s   & 1min12s & 1min2s \\ \hline
\end{tabular}
\end{table*}

\begin{figure*}[htbp]
    \centering
    \subfigure[]{
        \label{fig:subfig_a}
        \includegraphics[width=0.4\linewidth]{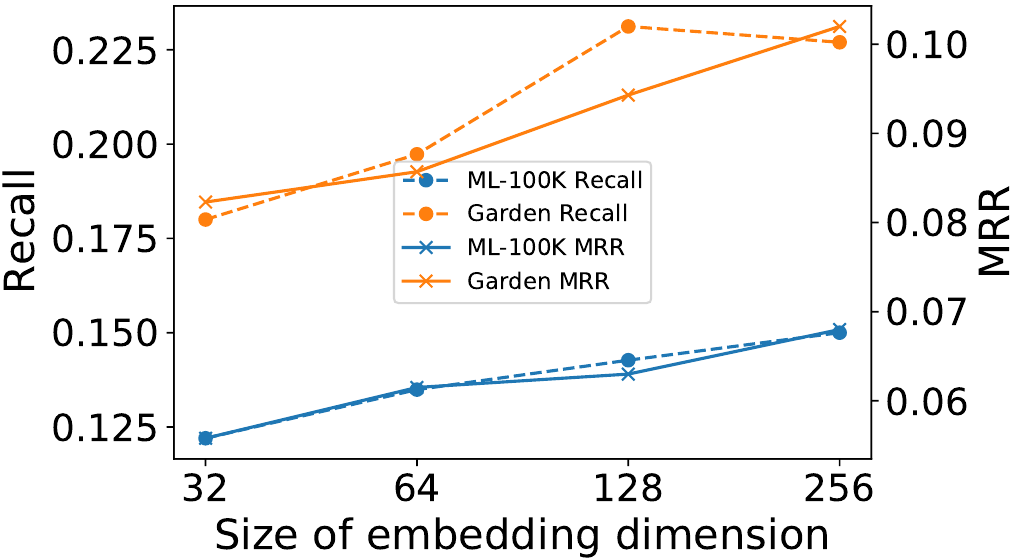}
    }
    \hfill
    \subfigure[]{
        \label{fig:subfig_b}
        \includegraphics[width=0.28\linewidth]{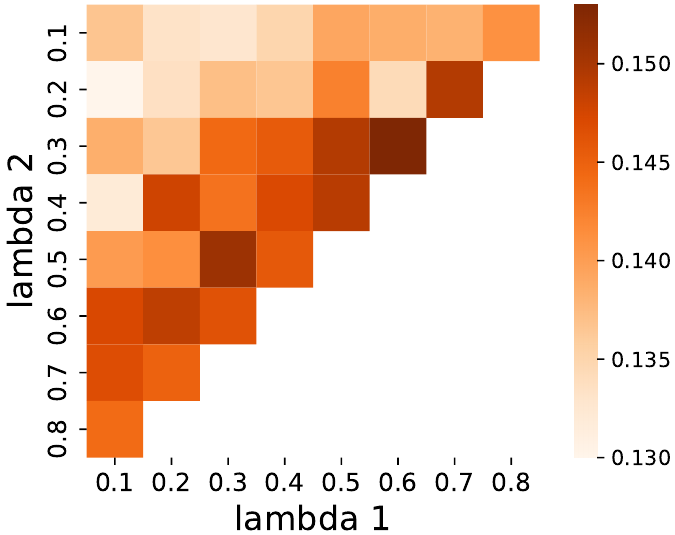}
    }
    \hfill
    \subfigure[]{
        \label{fig:subfig_c}
        \includegraphics[width=0.28\linewidth]{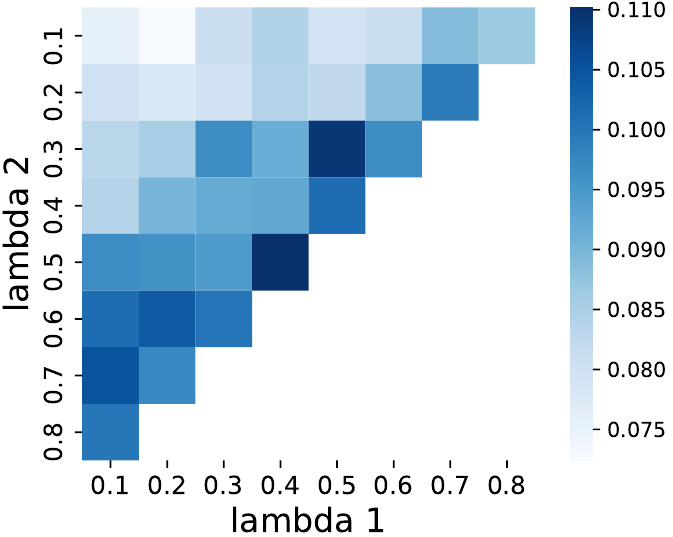}
    }
    \caption{Sensitivity analysis of \ours on three hyperparameters: \subref{fig:subfig_a} MRR and Recall@10 with different embedding size on ML-100K and Garden, \subref{fig:subfig_b} the heat-map of Recall@10 with different coefficients $\lambda_1$ and $\lambda_2$ on ML-100K, \subref{fig:subfig_c} the heat-map of Recall@10 with different coefficients $\lambda_1$ and $\lambda_2$ on Video.}
    \label{fig:subfig}
\end{figure*}

\subsubsection{Decoder outputs of interaction matrix and transition matrix}
We test if adding the decoder outputs of user-item interaction matrix and item transition matrix can improve the performance or not, compared with only using encoder hidden outputs. We add the reconstructed interaction matrix and transition matrix separately, and then add these two reconstructed matrices together. Specifically, we use a weighted sum to combine all the information (hidden outputs, reconstructed interaction matrix and transition matrix). As shown in Table \ref{tab:ablation1}, we observe that using all information is better than any other combination, confirming the usefulness of collaborative information and multi-hops transition information.

\subsection{Efficiency Comparison (RQ3)}
We use the future interaction prediction task to study the efficiency of JODIE, CoPE, FreeGEM and \ours. Other baselines like RRN show comparable efficiency with JODIE [14] and thus are omitted. FreeGEM requires multiple runs of offline/online SVD, JODIE and CoPE both run 50 epochs and choose the best performing model on the validation set as the optimal model. 

As shown in Table \ref{tab:running}, \ours is at least 16.3x faster than JODIE, at least 64.7x faster than CoPE, and at least 1.1x faster than the parameters-free method FreeGEM. This experiment confirms that \ours has higher computational efficiency than the compared methods, maily due to 1) the interaction matrix and transition matrix updates are incremental and 2) the encoder-decoder architecture in \ours is much more light-weight compared with RNNs (in JODIE) and GNNs (in CoPE).

\subsection{Sensitivity Analysis (RQ4)}

Here, we analyze how the key hyper-parameters affect the performance of \ours. We analyze the following hyper-parameters: the embedding size of the auto-encoder and the coefficients $\lambda_1$ and $\lambda_2$ for balancing the importance of different prediction scores in Equation \ref{con:coefficients}.

\subsubsection{Embedding Size.}
We analyze the impact of the embedding size in the autoencoder on the performance. To do this, we vary the embedding size of the autoencoder from 32 to 256 and calculate the MRR and Recall@10 scores for future interaction prediction on the Amazon Garden dataset and MovieLens ML-100K dataset. The trends on other datasets are similar. The results are shown in Figure \ref{fig:subfig} \subref{fig:subfig_a}. As the embedding dimension increases from 32 to 128, we find that the embedding size has a relatively positive effect on the performance of \ours, i.e., the performance of \ours consistently improve with increasing dimension size. This experiment demonstrates that \ours does not suffer from the common overfitting issue as other deep models.

\subsubsection{The Coefficients $\lambda_1$ and $\lambda_2$}
$\lambda_1$ controls the importance of collaborative information and $\lambda_2$ controls the importance of one-hop transition. The coefficients of collaborative information, one-hop transition and two-hops transition add up to one. Thus, once these two coefficients $\lambda_1$ and $\lambda_2$ are determined, the third one is immediately determined as well.
We observe that using a single type of information is not optimal. To make sure each source of information is used, we vary the coefficients from 0.1 to 0.8 and plot the heatmaps of Recall@10 on the ML-100K and Video datasets. As shown in Figure \ref{fig:subfig} \subref{fig:subfig_b} and Figure \ref{fig:subfig} \subref{fig:subfig_c},
the performance of \ours is not very sensitive to $\lambda_1$ and $\lambda_2$, i.e., we can observe near optimal performance with several different configurations.

\begin{figure*}
\begin{minipage}{0.62\linewidth}
    \centering
    \includegraphics[width=1\linewidth]{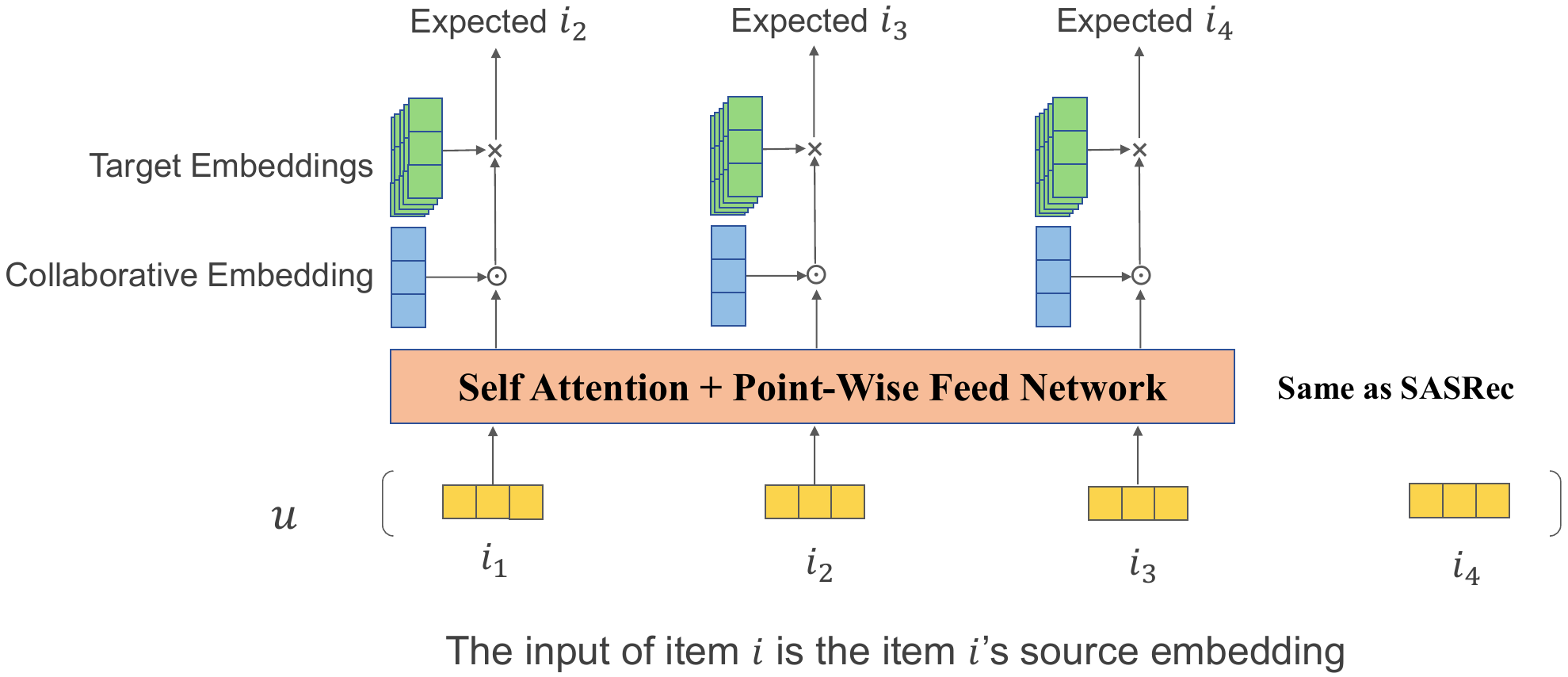}
    \caption{The process of integrating \ours with SASRec. Collaborative information embedding, source information embedding and target information embedding are  pre-trained by \ours.}\label{fig:case_study}
\end{minipage}
\hfill
\begin{minipage}{0.365\linewidth}
    \centering    \includegraphics[width=1\linewidth]{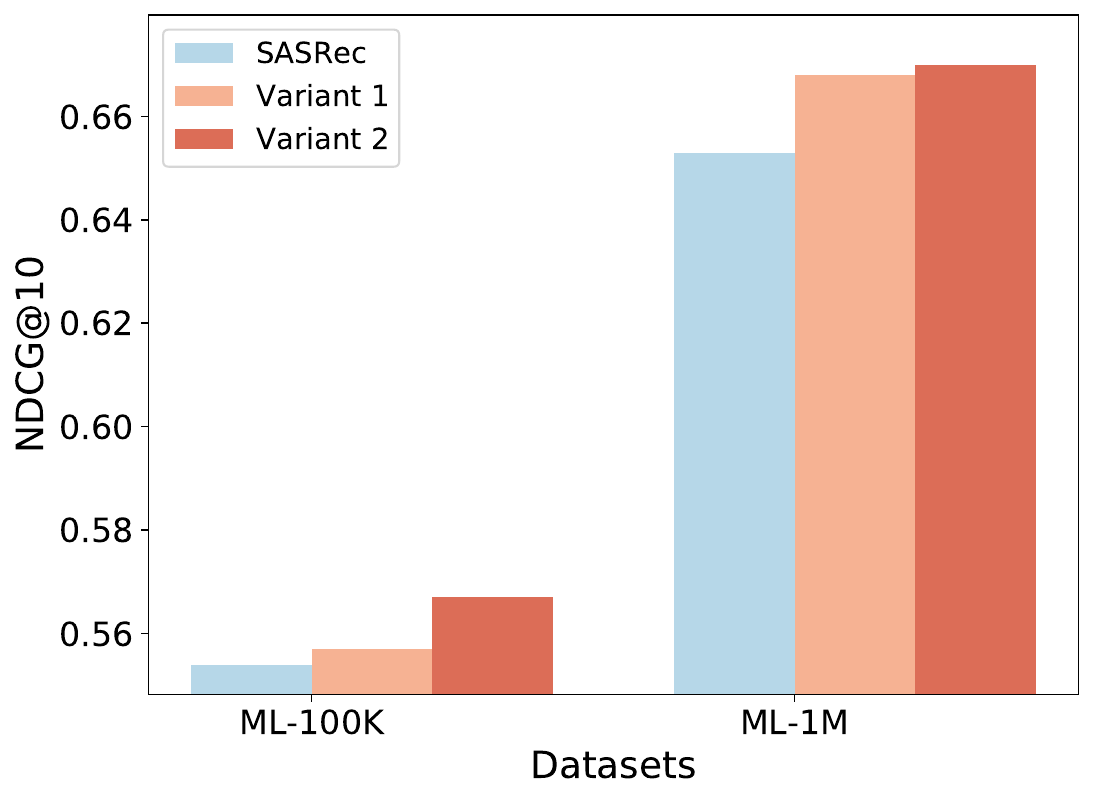}
    \caption{The recommendation accuracy of integrating \ours with SASRec on the two MovieLens datasets.}\label{fig:exp2_01}
\end{minipage}
\end{figure*}

\subsection{Integrating \ours with Other Sequential Recommendation Methods (RQ5)}

The major innovations in \ours, e.g., the embedding learned from multi-information encoder, the score functions (Equation \ref{con:Hadamard} and \ref{con:multiply}), and the combination of collaborative and transition information, are orthogonal to many existing sequential recommendation methods. Therefore, we believe \ours can be integrated with them to further improve their performance. Here, we take SASRec~\cite{kang2018self}, a well-known sequential recommendation method, as an example to illustrate how to integrate \ours with other sequential recommendation methods as shown in Figure \ref{fig:case_study}.

\subsubsection{Setting} 
We use the sequential recommendation task defined in SASRec~\cite{kang2018self}. 
All the data pre-processing and partitioning protocols are the same as SASRec: 1) we treat the presence of a review or rating as implicit feedback, 2) we use timestamps to determine the sequential order of actions, and 3) we split the historical sequence of each user into three parts: the most recent one interaction for test, the penultimate interaction for validation, and all remaining interactions for training.

\subsubsection{The Process of Integrating \ours with SASRec}
SASRec~\cite{kang2018self} is a sequential recommendation method based on the Transformer architecture~\cite{vaswani2017attention}.
We first briefly introduce its architecture.
Firstly, SASRec reads in the embedding sequence of items that a user has interacted with.
Then, after several Attention blocks (including self-attention layer and point-wise feed forward), we get the item embedding sequence after feature transformation.
Finally, the item embedding sequence after feature transformation will be input into the prediction layer (e.g., multi-layer perceptron), which will treat each item as a class, and train the model by optimizing the classification task.

While maintaining the main architecture (the Attention blocks) of SASRec, we use the collaborative information embedding $E_c$, source information embedding $E_s$ and target information embedding $E_t$, which are outputs from multi-information encoder, to improve the input and prediction layers of SASRec. Firstly, we will use the rows corresponding to source information embedding $E_s$ as input to SASRec, instead of the randomly initialized embedding.
Then, after passing through several Attention blocks, we will obtain the source information embedding after feature transformation.
Finally, in the prediction phase, we adopt a consistent approach with \ours.
Specifically, as Equation (\ref{con:Hadamard}), we first perform a Hadamard product between $E_c$ and the source information embedding after feature transformation to obtain the personalized transition embedding,
and then multiply personalized transition embedding with the target information embedding $E_t$ as Equation (\ref{con:multiply}) to further consider multi-hops transition information.

Similar to the ablation studies, we design two ablative variants for this study to further understand the effectiveness of our design. Specifically, we call that the model use collaborative information embedding as Variant 2, and  that without collaborative information embedding as Variant 1.

\subsubsection{Analysis}
We compare Variant 1 and Variant 2 with SASRec on two MovieLens datasets (ML-1M and ML-100k). We follow the evaluation strategy in prior works~\cite{he2017neural,koren2008factorization} to avoid heavy computation on all user-item pairs. For each user $u$, we randomly sample 50 negative items, and rank these items with the ground-truth item. The hidden dimension we use in our variants is 64. We use NDCG@10, which can simultaneously evaluate the ranking and recall of a prediction.

For more clear comparison, we plot the performance of SASRec and the two variants on the bar chart as shown in Figure \ref{fig:exp2_01}. We can observe that both variants are better than SASRec. The improvement of Variant 1 is 0.54\%-2.30\% in terms of NDCG@10 and the improvement of Variant 2 is 2.35\%-2.60\% in terms of NDCG@10. The experimental results indicate that multiplying the $i$-th row of the source information embedding by the target information embedding in Equation \ref{con:multiply} is helpful. 

Furthermore, compared with Variant 1, Variant 2 achieves relatively higher accuracy, i.e., higher NDCG@10 on both datasets. In other words, adding the operation, a Hadamard product on the collaborative information embedding of user $u$, is not only beneficial to the future interaction prediction task, but also beneficial to the sequential recommendation task defined in SASRec. Thus, we further validate the rationality of multiplying the $i$-th row of the source information embedding by the target information embedding, and operating a Hadamard product on the collaborative information embedding of user $u$.

From this case study, we can further conclude that 1) the embedding learned by \ours are effective for sequential recommendation, 2) our score functions are helpful for improving accuracy, and 3) it is beneficial to combine collaborative information with transition information for sequential recommendation tasks.

\section{Related Work}

In this section, we first introduce the  sequential recommendation methods, including recurrent neural network based methods, temporal point process based methods and graph based methods. Then, we introduce the autoencoder-based recommendation methods.

\subsection{Sequential Recommendation}
Many sequential recommendation methods are based on recurrent neural networks.
RRN~\cite{wu2017recurrent} models user and item interaction sequences with separate RNNs.
Time-LSTM~\cite{zhu2017next} adopts time gates to represent the time intervals.
LatentCross~\cite{beutel2018latent} incorporates contextual data into embeddings.
DeepCoevolve~\cite{dai2016deep} and JODIE~\cite{kumar2019predicting} generate node embeddings using two intertwined RNNs, and JODIE can also estimate user embeddings trajectories.

Another line of sequential recommendation work is based on temporal point processes.
Know-Evolve~\cite{trivedi2017know} and HTNE~\cite{zuo2018embedding} model interactions between users and items as multivariate point processes and Hawkes processes, respectively.
Shchur et al.~\cite{shchur2019intensity} directly model the conditional distribution of inter-event times, and therefore more accurately control the changes in user interests.
DSPP~\cite{cao2021deep} incorporates topology and long-term dependencies into the intensity function.
All these methods take time information into account by intensity functions.

Recently, more advancing graph based sequential recommenders are proposed.
TDIG-MPNN~\cite{chang2020continuous} captures both global and local information on the graph.
SDGNN~\cite{tian2021streaming} uses the state changes of neighbor nodes to learn the node embedding.
CoPE~\cite{zhang2021cope} uses an ordinary differential equation-based GNN to model the evolution of network.


\subsection{Autoencoder-based Collaborative Filtering}
Sedhain et al.~\cite{sedhain2015autorec} proposed AutoRec, the first autoencoder framework for collaborative filtering. Many researchers extend autoencoders to collaborative filtering for implicit feedback.
Liang et al.~\cite{liang2018variational} extended variational autoencoders to collaborative filtering for implicit feedback. Based on the Autorec and Transformer architectures, Dang et al.~\cite{dang2019attentional} proposed an attentional autoencoder for implicit recommendation. Zhu et al.~\cite{zhu2022mutually} proposed a mutually-regularized dual collaborative variational auto-encoder (MD-CVAE) for recommendation.

Moreover, some works incorporate content information to enhance the performance of autoencoder.
Li et al.~\cite{li2018deep} proposed a model that learns a shared feature space from heterogeneous data, leveraging heterogeneous auxiliary information to address the data sparsity problem of recommender systems. To solve the cold start problem, Lee et al.~\cite{lee2019basis} developed a hybrid CF technique incorporating CF with content information. For implicit trust relationships, Zeng et al.~\cite{zeng2021neural} proposed an autoencoder model based on implicit trust relationship between users, combining the intrinsic relationship of both the implicit trust information and user-item interaction behavior for collaborative recommendation.

\section{Conclusions}
Efficiently capturing changing interests in real-time recommendation is the key challenge faced by sequential recommendation methods.
This paper proposes \ours, which is an autoencoder-based incremental method for sequential recommendation.
\ours consists of an encoder and three decoders within an autoencoder architecture. The encoder takes as input the rows of the user-item interaction matrix, the rows of the item transition matrix, and the columns of the item transition matrix. By simultaneously capturing different types of information, the encoder generates informative representations of user interests and item transitions. The three decoders reconstruct these three types of information, enabling a comprehensive understanding of user preferences and short-term interests.
We conduct comprehensive experiments to verify the superiority of \ours, and the experimental results show that \ours can outperform the state-of-the-art
methods in accuracy while achieving significant improvement in efficiency.
In the future, we will try more complex encoder and decoder architectures, such as Transformers~\cite{kang2018self}.

\begin{acks}
This work was supported by the National Natural Science Foundation of China (NSFC) under Grants 62172106 and 61932007.
\end{acks}

\bibliographystyle{ACM-Reference-Format}
\balance
\bibliography{sample-base}

\appendix


\end{document}